\documentclass[aps,prb,showpacs,preprint]{revtex4-1}
\usepackage{amssymb}
\usepackage{amsmath}
\usepackage{graphicx}
\usepackage{dsfont}
\usepackage{times}

\begin{document}

\title{A Complex Network Analysis on The Eigenvalue Spectra of Random Spin Systems}
\author{Qiaomu Xue and Wenjia Rao}
\email{Correspondence, wjrao@hdu.edu.cn}
\affiliation{School of Science, Hangzhou Dianzi University, Hangzhou 310027, China.}
\date{\today }

\begin{abstract}
Recent works have established a novel viewpoint that treats the eigenvalue
spectra of disordered quantum systems as time-series, and corresponding
algorithm such as singular-value-decomposition has proven its
advantage in studying subtle physical quantities like Thouless energy and non-ergodic
extended regime. On the other hand, algorithms from complex networks have long been known as
a powerful tool to study highly nonlinear time-series. In this work, we combine these two
ideas together. Using the particular algorithm called visibility graph (VG)
that transforms the eigenvalue spectra of a random spin system into complex
networks, it is shown that the degree distribution of the resulting network is
capable of signaturing the eigenvalue evolution during the thermal to
many-body localization transition, and the networks in the thermal
phase have a small-world structure. We further show these results are
robust even when the eigenvalues are incomplete with missing levels,
which reveals the advantage of the VG algorithm.
\end{abstract}

\maketitle

\section{Introduction}

The thermal and many-body localized (MBL) phases, as the two generic phases
in isolated quantum systems, have attracted much attention in recent
condensed matter society. A thermal phase acts as the heat bath for its
own subsystem, which results in an extensive quantum entanglement that follows
volume-law. In contrast, a MBL phase is where localization persists in the
presence of many-body interactions, which leads to small (area-law) entanglement.
The different scaling behaviors of entanglement provide the modern
understanding about these two phases\cite%
{Kjall2014,Yang2015,Serbyn16,Maksym2015,Kim,Bardarson,Abanin,Garcia}.

More traditionally, the standard understanding about these phases relies
on the statistical description of their eigenvalue spectra, whose
mathematical foundation is provided by the random matrix theory (RMT)\cite%
{Mehta,Haake}. Specifically, a thermal phase has correlated eigenvalue spectra that follow
the Wigner-Dyson ensemble, which further divides into three classes
depending on the system's symmetry; in contrast, the MBL phase has
uncorrelated spectra that follow the Poisson ensemble\cite%
{Oganesyan,Avishai2002,Regnault16,Regnault162,Huse1,Huse2,Huse3,Luitz,Serbyn,Sierant19,Sierant20,SierantPRL,Garcia2006}.

Despite the conventional RMT-based descriptions of the eigenvalue
spectra such as the level spacing distributions or $\Sigma^2$ statistics,
researchers have always been searching for new methods that go beyond RMT.
One such approach, is to view the eigenvalue spectra $\{E_i\}$ of
random matrix models as \emph{time series}, where the eigen-level index $i$ plays
the role of virtual time. This idea was originally proposed numerically two decades
ago\cite{TimeSeries2002} and then proved analytically using the power-spectrum function analysis\cite{Faleiro,Kan,Molina}. Heuristically, this idea is quite natural because the eigenvalue
$E_i$ of a random system necessarily becomes a stochastic variable that follows
certain probability distribution, and the eigenvalue correlations can be translated
into the history dependence in normal time series. Later, the method of singular-value-decomposition
(SVD) was employed to decompose the eigenvalue spectrum into trend and fluctuating modes, which
allows us to unfold the spectrum in a data-adaptive manner\cite{Vargas1}
that avoids the ambiguity raised by conventional unfolding procedure\cite{Gomez2002}.
Taking this advantage, combined with mode analysis, we are able to evaluate the spectral
(non-)ergodicity of a random matrix model\cite{Jackson,add1,add2,add3}. Moreover, the scaling
behavior of the singular values after SVD (so-called scree plots) was shown to follow a
power-law distribution that reflects the ergodicity property of the system, which was then
successfully applied to a number of random matrix models\cite{Vargas2,Vargas3}. More recently,
it is shown that the scree plots can be used to identify the non-ergodic extended regime
in the Rosenzweig-Porter model\cite{Khaymovich,Berkovits1} and Anderson model\cite{Berkovits2},
which is afterward generalized to related physical and random matrix models\cite{Berkovits3,Berkovits4,Berkovits5,Rao22,Ni,Rao23}.

On the other hand, the methods of complex network (CN), have long
been known as a powerful tool in studying time series, especially the ones
with high non-linearity\cite{Zou2019,Zhang2006,ts1,ts2,ts3,ts4}. However, to the best of our knowledge,
CN algorithms have not been used to study the eigenvalue spectra of disordered
quantum systems so far. In this work, we will try to fill in this gap.
Specifically, we employ a CN algorithm called visibility graph (VG)\cite{vg1,vg2} to
transform the eigenvalue spectra of random spin systems into CNs, and show
the topological characterizations of the resulting CNs can faithfully reveal the
eigenvalue evolution during the thermal-MBL transition. Notably, we find the
CNs in the thermal phase has a small-world\cite{Nat1998} structure, which persists
to the transition region. As a result,
the thermal, MBL and transition region can be distinguished from the perspective of
CNs. Furthermore, given the topological nature of the VG
algorithm, we show that this method works even in cases of incomplete eigenvalue
spectra with missing levels, which reveals its advantage. It is also interesting
to note that RMT has been employed to study the properties of complex
networks (e.g. by analyzing the adjacent matrices of the latter)\cite{Jalan1,Jalan2},
while in this work we are taking the opposite direction.

This paper is organized as follows. In Sec.\ref{sec2} we introduce the
target physical model and the VG algorithm. In Sec.\ref{sec3} we study the
degree distribution of the resulting networks, and discuss its relation to
normal level spacing distribution. In Sec.\ref{sec4} we discuss the
clustering and shortest-path-lengths of the networks, and show the VG
in the thermal phase has a small-world structure. In Sec.\ref{sec5} we show these
results are robust in cases with incomplete eigenvalue spectra, which reveals the
algorithm's advantage. Discussion and conclusion are given in Sec.\ref{sec6}.

\section{Visibility Graph Algorithm}

\label{sec2} \bigskip

As the test ground of this work, we consider the canonical model in the area
of MBL physics: the one-dimensional spin-1/2 chain with random magnetic
fields, whose Hamiltonian is%
\begin{equation}
H=\sum_{i=1}^{L}\mathbf{S}_{i}\cdot \mathbf{S}_{i+1}+h\sum_{i=1}^{L}\sum_{%
\alpha =x,z}\varepsilon _{i}^{\alpha }S_{i}^{\alpha }\text{,}  \label{equ:H}
\end{equation}%
where the anti-ferromagnetic coupling strength is set to be unity, and $%
\varepsilon _{i}^{\alpha }$s are random variables in $\left[ -1,1\right] $, $%
h$ is referred as the randomness strength. This Hamiltonian undergoes an
thermal-MBL transition at $h_{c}\simeq 3$, with the eigenvalue statistics
evolving from Gaussian Orthogonal Ensemble (GOE) to Poisson\cite{Regnault16,Regnault162,Rao21}.
Compared to the more widely-studied model with $\varepsilon _{i}^{x}=0$, this model breaks
total $S_{z}$ conservation and is less affected by the finite-size effect\cite{Rao21}.
In this work, we choose to simulate an $L=13$ system, with the local Hilbert space
dimension being $2^{13}=8192$. We exactly diagonalize $H$ to obtain the
eigenvalue spectra $\left\{ E_{i}\right\} $\ at several $h$s, and at each $%
h $ we generate $N_{s}=1000$ samples. To avoid the influence by the potential
existence of many-body mobility edges, we select only the middle $N_{d}=2000$ levels for
later complex-network investigations. Such a model has been studied with the
SVD method in an earlier work\cite{Rao22}, which allows for a direct comparison.

Among various complex network algorithms that turn time series into
networks, we employ the visibility graph (VG) algorithm in this work\cite{vg1,vg2}, whose
construction goes as follows. Given an eigenvalue spectrum $\left\{ E_{i}\right\} $,
where $i$ is the level index and now treated as the virtual time index, we draw
$\left\{ E_{i}\right\} $ consecutively as $\left( i,E_{i}\right) $ in the two-dimensional space,
which are called \emph{vertices} in the VG terminology. Two vertices $\left(
i,E_{i}\right) $ and $\left( j,E_{j}\right) $ are connected by an \emph{edge}
if the criteria%
\begin{equation}
\frac{E_{j}-E_{i}}{j-i}>\frac{E_{j}-E_{k}}{j-k}
\end{equation}%
if fulfilled for all $k\in \left[ i,j\right] $. In other words, two vertices
are connected if there is no internal vertex that blocks the line of sight
from $\left( i,E_{i}\right) $ to $\left( j,E_{j}\right) $, which explains
the algorithm's name. As an eigenvalue spectrum of a Hamiltonian, the
$\left\{ E_{i}\right\} $ has a trivial trend that $E_{j}>E_{i}$ if $j>i$, so
this VG algorithm (called natural VG) would be more appropriate than other
kinds of VG such as the Horizontal Visibility Graphs (HVG)\cite{HVG}.

The collections of all vertices and edges comprise the visibility graph.
Here we focus on the simplest case that all the edges are unweighted and
undirected. Clearly, in this setting, all vertices are connected to their
neighboring ones. For a demonstration, we take out one typical sample of $%
\left\{ E_{i}\right\} $ in the $h=1$ dataset, and draw the behavior of the
middle $20$ eigenvalues in Fig.1(a), where the grey lines are the connected
edges under the VG criteria. It's worth noting that the way to draw the VG
is not unique, for example, Fig.1(b) displays another configuration which is
equivalent to Fig.1(a) in a complex-network sense. Therefore, the VG algorithm
is of topological nature, and in the following we will calculate several typical
topological characterizations of the VGs and explore their relation with the eigenvalue statistics.

\begin{figure}[t]
\centering
\includegraphics[width=\columnwidth]{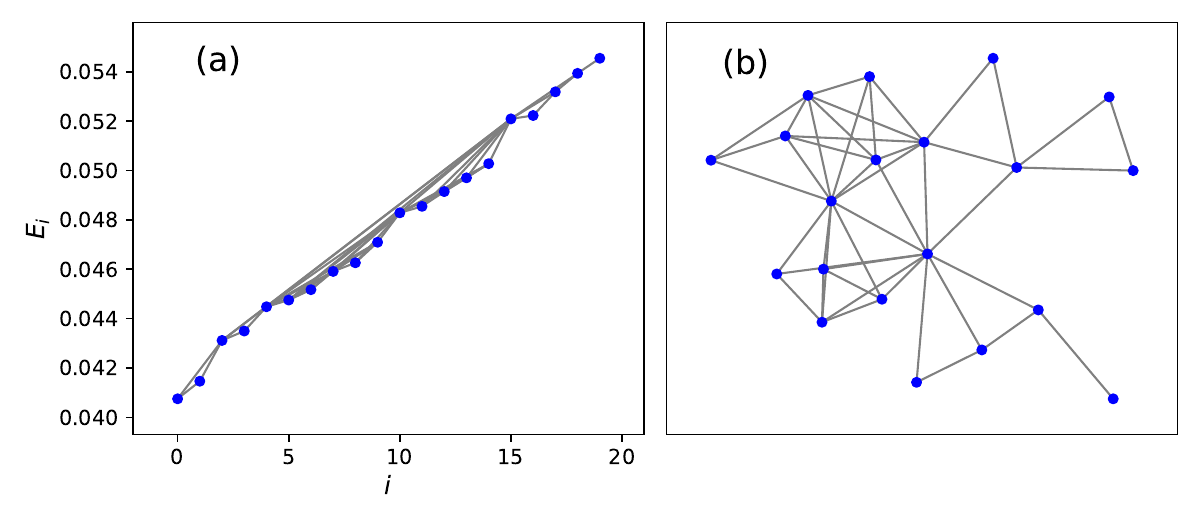}
\caption{(a) The behavior of a segment $E_i$, where the grey lines corresponds to the connected edges
under the VG criteria. (b) A topologically equivalent network configuration of (a).}
\label{fig:1}
\end{figure}

\section{Degree Distributions}

\label{sec3}

Among various topological characterizations of a complex network, the most
important one is the distribution of its vertices' degrees. A \emph{degree}
of a vertex $\left( i,E_{i}\right) $, denoted by $k_{i}$, is defined as the
number of vertices that are connected with $\left( i,E_{i}\right) $ under
the VG criteria. By VG's construction we know that each vertex (except for the
starting and ending ones) is at least connected to its two neighbors, i.e. $%
k_{i}\geq 2$. We are most interested in the probability distribution $%
P\left( k\right) $ and its mean value $\langle k\rangle $. For a
demonstration, we take out one sample from the dataset with $h=1$ and $h=5$,
which belong to the thermal and MBL phase respectively, the corresponding $%
P\left( k\right)$ are drawn in Fig.2(a).

\begin{figure}[t]
\centering
\includegraphics[width=\columnwidth]{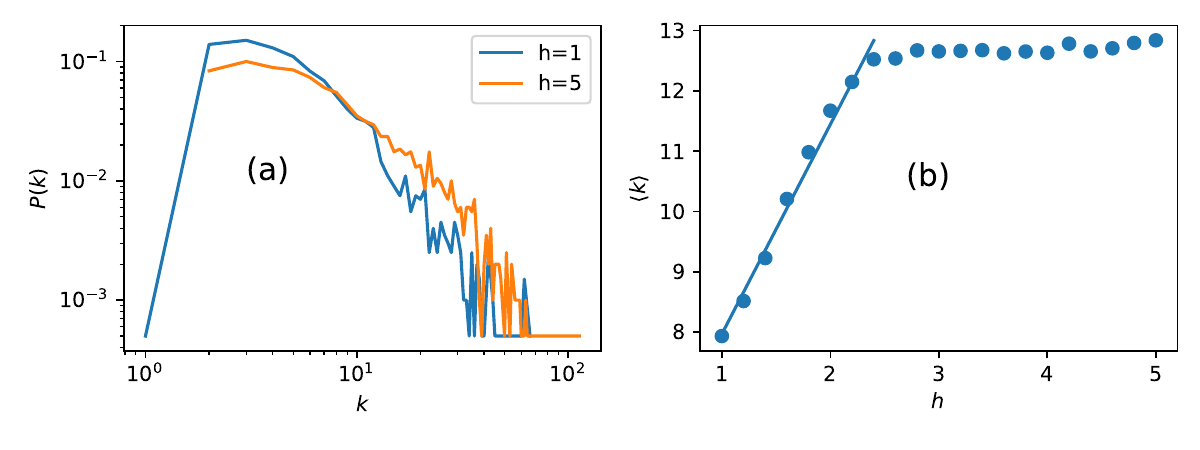}
\caption{(a) The degree distribution $P(k)$ of a typical spectrum in the thermal ($h=1$) and
MBL ($h=5$) phase, the latter has significantly larger $k$. (b) $\langle k\rangle$ as a function
of randomness strength $h$.}
\label{fig:2}
\end{figure}

As we can see, the VG of the MBL phase has significantly larger values of $k$
than that of the thermal phase, which indicates the larger eigenvalue
repulsion will results in smaller edge degrees. This hints us the mean value
of the edge degree $\langle k\rangle $ may be used to signature the
eigenvalue evolution during the thermal-MBL transition. We therefore compute
$\langle k\rangle $ among all spectra samples, and
plot them as a function of the randomness strength $h$ in Fig.2(b). Interestingly,
we find that $\langle k\rangle $ grows linearly with the randomness strength $h$ in the thermal
phase, whose formula is fitted to be $\langle k\rangle \simeq 3.48h+4.49$.
While $\langle k\rangle $ saturates to a value $\langle k\rangle \simeq 12.8$
at $h\simeq 2.6$, which is close to the thermal-MBL transition point $%
h_{c}\simeq 3$\cite{Regnault16,Regnault162,Rao21}. This means $\langle k\rangle $
is capable of reflecting the eigenvalue evolution during the phase transition,
which is the first non-trivial finding of this work.

To further explore the implication of $\langle k\rangle $ on eigenvalue
statistics, recall the normal way to characterize the eigenvalue statistics
in random matrix theory is the probability distribution of the level
spacings $\left\{ s_{i}=E_{i+1}-E_{i}\right\} $, which follows the GOE
distribution $P\left( s\right) =\frac{\pi }{2}se^{-\pi s^{2}/4}$ in the
thermal phase and the Poisson distribution $P(s)=e^{-s}$ in the MBL phase,
both of which are comprised a polynomial term that reflects the level
repulsion (or equivalently the system's symmetry) and an exponential term
that reflects the large-$s$ decaying (or the level rigidity). This reminds us
the widely-accepted formula for the $P\left( s\right) $ evolution during
thermal-MBL transition proposed by Serbyn and Moore\cite{Serbyn}, i.e.%
\begin{equation}
P\left( \beta _{1},\beta _{2},s\right) =C_{1}s^{\beta _{1}}\exp \left(
-C_{2}s^{2-\beta _{2}}\right) \text{.}  \label{equ:dist}
\end{equation}%
where \thinspace $C_{1},C_{2}$\ are normalization parameters determined by $%
\int P\left( s\right) ds=\int sP\left( s\right) ds=1$. For case deep in the
thermal phase, we have $\beta _{1}=1,\beta _{2}=0$ standing for the GOE
distribution; and for case deep in the MBL phase, $\beta _{1}=0,\beta _{2}=1$
corresponding to the Poisson ensemble. Ref.[\onlinecite{Serbyn}] shows that the
exponential parameters $\beta _{2}$ evolves faster than the polynomial
term, i.e. $\beta _{2}$ saturates to $1$ before the transition happens, which
is attributed to the Griffiths regime near the transition point. Since we also
find the saturation of $\langle k\rangle $ happens earlier than the MBL
transition in Fig.2(b), we are led to propose that $\langle k\rangle $ \emph{only}
reflects the exponential term of $P(s)$, while is incapable of detecting the
short-range level repulsion term, in other words, it is insensitive to the
system's symmetry.

To verify above idea, we conduct two comparative tests. The first one is the
the Gaussian $\beta $ ensemble which generalizes the conventional Wigner
classes with $\beta =1,2,4$ to arbitrary positive value of $\beta $, and hence $%
P\left(\beta,s\right) \sim s^{\beta }e^{-C\left( \beta \right) s^{2}}$. If our conjecture holds,
all $\beta$ ensembles should have identical $\langle k\rangle$ in their VGs. The
eigenvalue spectra of $\beta $ ensemble can be obtained by diagonalizing the
following tridiagonal ``parent'' matrix\cite%
{Beta}%
\begin{equation}
M_{\beta }=\frac{1}{\sqrt{2}}\left(
\begin{array}{ccccc}
x_{1} & y_{1} &  &  &  \\
y_{1} & x_{2} & y_{2} &  &  \\
&
\begin{array}{ccc}
\text{.} &  &  \\
& \text{.} &  \\
&  & \text{.}%
\end{array}
&
\begin{array}{ccc}
\text{.} &  &  \\
& \text{.} &  \\
&  & \text{.}%
\end{array}
&
\begin{array}{ccc}
\text{.} &  &  \\
& \text{.} &  \\
&  & \text{.}%
\end{array}
&  \\
&  & y_{N-2} & x_{N-1} & y_{N-1} \\
&  &  & y_{N-1} & x_{N}%
\end{array}%
\right)  \label{equ:Beta}
\end{equation}%
where the diagonals $x_{i}\,$($i=1,2,...,N$) follow the normal distribution $%
\mathit{N}\left( 0,2\right) $, and $y_{k}$ ($k=1,2,...,N-1$) follows the $%
\chi $ distribution with parameter $\left( N-k\right) \beta $. When $\beta
=1 $, it reduces to the standard GOE. Without loss of generality, we select
two non-standard values of $\beta =0.4,0.8$, together with the standard GOE ($\beta=1$),
and compute the corresponding $\langle k\rangle $ after transforming them into VGs,
and the results are given in Table I, as expected, they are quite close.

For the second model, we consider the short-range plasma model (SRPM) that frequently
appears in the study of intermediate random matrix ensembles\cite{SRPM},
which is widely accepted as the critical distribution at the MBL transition
point. SRPM treats the eigenvalues spectrum as an ensemble of
one-dimensional particles with only nearest neighboring logarithmic
interactions, and it is known\cite{SRPM} that the large $s$ behavior of its level
spacing follows $P\left( s\right) \sim s^{\beta }e^{-\left( \beta +1\right)
s}$ where $\beta $ is the Dyson index controlling the strength of level
repulsion. It is well-established that the eigenvalue spectrum of SRPM with
index $\beta $ is identical to the spectrum comprised of every $\left( \beta
+1\right) $-th eigenvalue from the Poisson ensemble\cite{Daisy}, which
enables us to effectively obtain SRPM's eigenvalue spectra to do VG
analysis. If our conjecture is corrected, the $\langle k\rangle $ of SRPM
with different $\beta$ should all be close to the Poisson ensemble.
We select $\beta=1,2$ as representative examples, and confirm this conjecture,
as shown in Table I.

\begin{table}
\begin{tabular}{|l|l|l|l|l|l|l|}
\hline
Model & $\beta =0.4$ & $\beta =0.8$ & $\beta =1$(GOE) & Poisson & SRPM$%
\left( \beta =1\right) $ & SRPM$\left( \beta =2\right) $ \\ \hline
$\langle k\rangle $ & 8.19 & 8.14 & 8.23 & 14.49 & 14.55 & 14.60 \\ \hline
\end{tabular}%
\caption{$\langle k\rangle$ in several random matrix ensembles. The first
three are the Gaussian $\beta$ ensembles, whose level spacing distribution decays
as $e^{-s^2}$; the rest are Poisson/SRPM, with $P(s)$ decays as $e^{-s}$.}
\end{table}

To conclude, we have shown that $\langle k\rangle $ is able to reflect the evolution
of the exponential term of the level spacing distribution -- which is a
long-range level correlation, while it is insensitive to the short-range
level repulsion parameter, which results in an under-estimation for the
thermal-MBL transition point. In fact, similar results have been reported
using the SVD method in Ref.[\onlinecite{Rao22}], which together justifies the idea that
treats the eigenvalue spectra as time series. To further explore the usefulness
of the VG algorithm, we are led to the next section.

\section{Clustering and shortest-path-lengths}

\label{sec4}

The next crucial characterization of a VG is the \emph{clustering coefficient%
} $c_{i}$ of a vertex, which is defined as%
\begin{equation}
c_{i}=\frac{D_{i}}{k_{i}\left( k_{i}-1\right) /2}
\end{equation}%
where $k_{i}$ is the degree of vertex $i$, and $D_{i}$ is the number of
edges between these $k_{i}$ vertices that are connected to $i$. Clearly, the
maximum value of $D_{i}$ is $k_{i}\left( k_{i}-1\right) /2$,
and so $c_{i}\in \lbrack 0,1]$. The value of $c_{i}$ measures how close the $%
i$-th vertex's connected vertices are clustered together, which explains its
name.

We compute the average value of the clustering coefficients $\langle c\rangle$ of the VGs
generated from eigenvalue spectra in different randomness strengths. Unlike the
degree distribution in previous section, we do not find significant differences in $\langle c\rangle$
between the thermal and MBL phases. Specifically, for VGs from $h=1,2,3,4,5$, we
find $\langle c\rangle = 0.74,0.72,0.71,0.71,0.71$ respectively.

Above result means the clustering is not a quantity that is sensitive to
eigenvalue evolution, but this is not the end of the story. In the context
of complex network, a densely clustered network could be a regular network
or a ``small world'' (SW) network\cite{Sci1999,Nat1998}. The SW network
is signatured by the existence of ``short cuts'' in a
network, typical examples include the social network and neural network.
To identify an SW network, we need to study the \emph{shortest-path-lengths}.
In a complex network, two disconnected vertices $i$ and $j$ can be connected by
several intermediate edges, which is called a \emph{path}, its length is defined as the number of these
internal edges. The path between two vertices are generally not unique, and
we are most interested in the shortest one. In an SW network, the mean
shortest-path-length $L$ grows logarithmically with the network's size $N$,
that is%
\begin{equation}
L\left( N\right) \sim \ln N.
\end{equation}%
for large $N$.

We therefore calculated the $L\left( N\right) $ for the VGs of the physical
eigenvalue spectra, and find that $L(N)$ in the thermal phase has
clear $\ln N$ behavior, which maintains even to the transition
region; while for MBL phase $L\left( N\right) $ will saturates to
a finite value, as demonstrated in Fig.3, where we select $h=1,3,5$ to
represent the thermal phase, transition region and MBL phase respectively.
Therefore, combined with the large clustering coefficient, we conclude the
VGs in thermal and transition region have SW structures.

\begin{figure}[t]
\centering
\includegraphics[width=0.6\columnwidth]{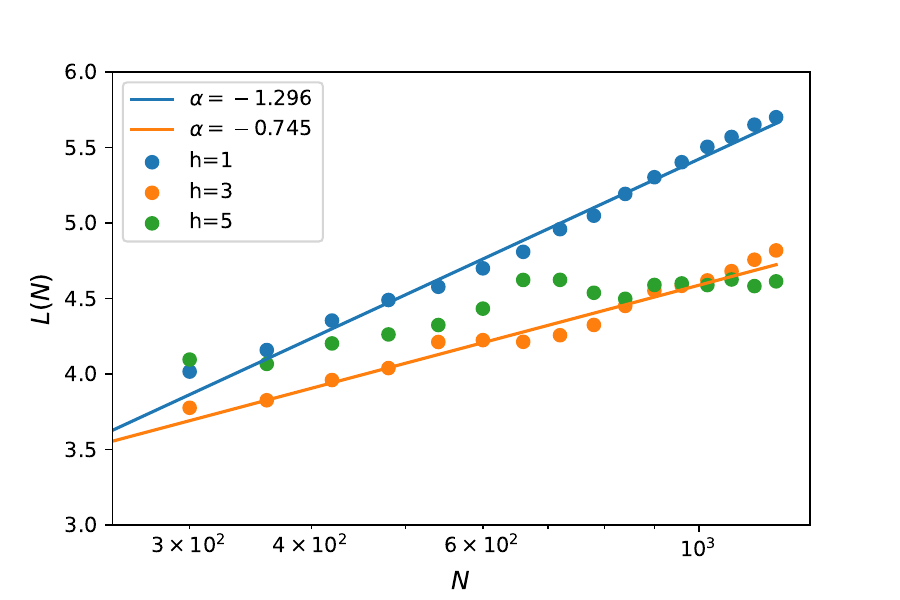}
\caption{The mean shortest-path-length $L(N$) as a function of $N$, where the x-axis
is plotted using a log scale. Clear small-world structure
$L(N)\sim ln(N)$ appears in the thermal phase ($h=1$), which maintains even to the transition region ($h=3$).}
\label{fig:3}
\end{figure}

To this stage, we are able to use the VG characterizations to distinguish the
thermal, MBL phases and transition region, as sketched in Table II. This
justifies the basic usefulness of the VG algorithm.

\begin{table}
\begin{tabular}{|l|l|l|l|}
\hline
& $\text{Thermal}$ & $\text{Transition}$ & $\text{MBL}$ \\ \hline
$\langle k\rangle $ & $\text{Small}$ & $\text{Large}$ & $\text{Large}$ \\
\hline
$\text{SW}$ & $\text{Yes}$ & $\text{Yes}$ & $\text{No}$ \\ \hline
\end{tabular}%
\caption{The qualitative criteria to distinguish thermal, MBL phases and
transition region using VG's topological characterizations.}
\end{table}

\section{Cases with incomplete eigenvalue spectra}

\label{sec5}

Despite the ability to distinguish phases, to make this work worthwhile,
it's still questionable about VG's advantage over traditional approaches. To
verify this, we consider the case that the eigenvalue spectra are incomplete
with missing levels. The situation of missing levels is common from the
experimental aspect, particularly in the area of heavy nuclei, which is
exactly the initial playground of random matrix theory back in the 1960s.
Analytical treatment for the cases with missing levels can be performed via
high-order level spacing distributions, as studied in Ref.[\onlinecite{Bohigas}].
However, since the VG is by construction a topological object, as demonstrated in Fig.1,
it is hopeful that the topological characterizations of VGs are robust to spectra with
missing levels, at least when the missing probability is not too larger.

To simulate an incomplete spectrum, we set a quantity $p$ that
is the probability a level is missing in a spectrum, and so we will end up
with a spectrum with mean length $N_{d}*(1-p)$. We then turn the incomplete
spectra into VGs, and explore whether the characterizations of the latter are
similar to the case with complete spectra.

\begin{figure}[t]
\centering
\includegraphics[width=0.9\columnwidth]{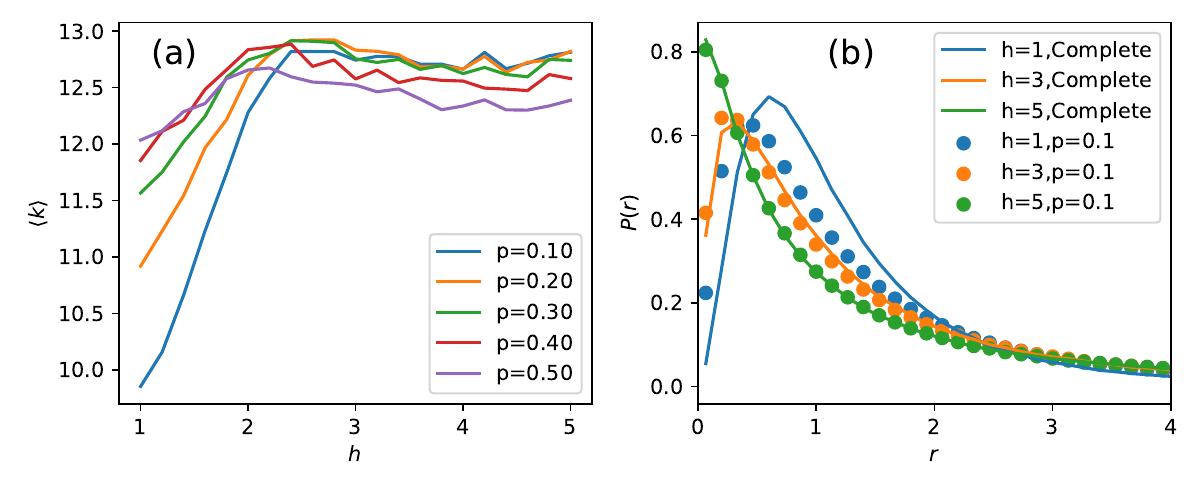}
\caption{(a) The evolution of $\langle k\rangle$ as a function of randomness strength $h$ in
spectra with missing levels, where the probability of missing a level is $p$. The evolution
resembles the case with complete spectra in Fig.2(b) up to $p=0.4$. While when $p$ approaches $0.5$,
the VG structure breaks down. (b) The normal spacing ratio distribution of the complete spectra
and incomplete spectra with $p=0.1$ at the ergodic ($h=1$), MBL ($h=5$) and transition region ($h=3$).}
\label{fig:4}
\end{figure}

As it turns out, the mean degree $\langle k\rangle$ is robust. In
Fig.4(a), we draw the evolution of $\langle k\rangle$ as a
function of randomness strength $h$ in cases with level missing probability $%
p\in[0.1,0.5]$. We see the tendency of $\langle k\rangle$ is similar to the
complete spectra case (that is, $\langle k\rangle$ grows with increasing $h$,
and saturates before the transition point) even when $p\sim 0.4$, which is a fairly large value.
When $p$ grows to $0.5$, the number of missing levels is close to the kept
levels, and hence the structure of VG breaks down, and $\langle k\rangle$
becomes fluctuating.

For a heuristic comparison, we also computed the commonly studied spacing ratio
distribution $P(r)$, where $\left\{r_i=\frac{E_{i+2}-E_{i+1}}{E_{i+1}-E_{i}}\right\}$
in the complete spectra and incomplete spectra with a small $p=0.1$, the results
are displayed together in Fig.4(b), where we choose $h=1,5,3$ to represent the thermal,
MBL and transition region respectively. As we can see, the deviations between complete
and incomplete spectra are larger in the thermal phase than in the MBL phase, which is
expected since the eigenvalue correlations in the former are larger, hence the
effect of missing levels will be more severe. In fact, we find that $P(r)$
of the incomplete spectra in the thermal phase ($h=1$) is quite close to the transition
area ($h=3$), which destroys the possibility to identify phases from $P(r)$. This situation
becomes more severe when missing probability grows. These results,
together with Fig.4(a), further justifies the advantage of the VG algorithm.

On the other hand, the tests for the mean shortest-path length $L(N)$ do not
give conclusive results. This is because the length of incomplete spectrum is not
large enough, so the dependence of $L(N)$ on $N$ is ambiguous between polynomial
and logarithmic. To improve the result we need to obtain more spectra samples
with larger system sizes, which is unaccessible with our current computational
resources. Nevertheless, the results for $\langle k\rangle$ in Fig.4 have verified
the VG's applicability in the case with missing levels.

\section{Conclusion and Discussion}

\label{sec6}

With the central idea that treats the eigenvalue spectra as multi-variant time series,
in this work we employed the visibility graph (VG) algorithm to make a
complex network analysis on the eigenvalues of random spin systems.
The main findings are as follows.

First of all, we find the degree distribution of the VGs generated from eigenvalue
spectra can faithfully reflects the evolution of eigenvalue statistics, specifically,
spectra with larger correlation in the thermal phase have significantly smaller degrees
than the MBL phase with uncorrelated spectra. We also verified the mean degree $\langle k\rangle$
is closely related to the exponential term in the level spacing distribution, while is insensitive
to the short-range level repulsion, or the system's symmetry. Second, we find that the
spectra in the thermal phase have a small-world structure, which
maintain even to the transition region. Thus, the degree distribution and
SW structure can be utilized together to distinguish the thermal, MBL and transition regions.
Last but not least, we show the VG algorithm is applicable in cases with incomplete
spectra, which stems from VG's topological construction and reveals the advantage
of the complex network methodologies.

It is worth noting that the VG employed in this work is one of the simplest complex
network configurations, while it is already capable of providing non-trivial
physical information, which justifies the validity of employing complex networks
to study the disordered quantum systems. Present paper serves as a
preliminary work in this direction that is by no means complete, there is certainly many future
directions to explore. To name just a few here. First of all, there are many other
network characteristics that do not appear in this work, such as the adjacent
matrix and assortive mixing, their relation to the eigenvalue correlations
deserves to be explored. Second, the VG employed in this work is insensitive
to the system's symmetry, which may attribute to its simple construction,
its tempting to bring some sophistication to VGs, for instance by adding weights
and directions to the edges. Third, complex networks naturally captures the
long-range level correlations in a spectrum, hence it's hopefully to study subtle
physical quantities like Thouless energy and Grifiths regime, during which the
unfolding procedure can be avoided. This necessarily requires larger datasets,
and will be studied in a future work.

\section{Acknowledgements}

This work is supported by the Zhejiang Provincial Natural Science Foundation
of China under Grant No.LY23A050003.

Data Availibility Statement: The data that support the figures within this
paper are available from the corresponding author upon reasonable request.

\end{document}